\documentclass{jfm}
\usepackage{graphicx}
\usepackage{epstopdf, epsfig}

\usepackage{amssymb}
\usepackage{subfigure}
\usepackage{pifont}

\shorttitle{Dissolution instability and roughening transition}
\shortauthor{P. Claudin, O. Dur\'an and B. Andreotti}

\title{Dissolution instability and roughening transition}

\author{
Philippe Claudin\aff{1}
\corresp{\email{philippe.claudin@espci.fr}},
\and Orencio Dur\'an \aff{2}
\and Bruno Andreotti\aff{3}
}

\affiliation{
\aff{1}Physique et M\'ecanique des Milieux H\'et\'erog\`enes, UMR 7636 ESPCI -- CNRS -- Univ.~Paris-Diderot -- Univ.~P.~M.~Curie, 10 rue Vauquelin, 75005 Paris, France.
\aff{2}Dept. of Ocean Engineering, Texas A \& M Univ., College Station, TX 77843-3136, USA.
\aff{3}Laboratoire de Physique Statistique, UMR 8550 Ecole Normale Sup\'erieure -- CNRS -- Univ.~Paris-Diderot -- Univ.~P.~M.~Curie, 24 rue Lhomond, 75005 Paris, France.
}

\newcommand{\dx}{\partial_x}
\newcommand{\dz}{\partial_z}

\def\strutdepth{\dp\strutbox}
\def\nw#1{\strut\vadjust{\kern-\strutdepth\vtop to0pt{\vss\hbox to\hsize
{\hskip\hsize\hskip5pt$\leftarrow$\hss\strut}}}{\em #1}}

\usepackage[numbers]{natbib}

\pdfoutput=1

\begin{document}

\maketitle

\begin{abstract}
We theoretically investigate the pattern formation observed when a fluid flows over a solid substrate that can dissolve or melt. We use a turbulent mixing description that includes the effect of the bed roughness. We show that the dissolution instability at the origin of the pattern is associated with an anomaly at the transition from a laminar to a turbulent hydrodynamic response with respect to the bed elevation. This anomaly, and therefore the instability, disappears when the bed becomes hydrodynamically rough, above a threshold roughness-based Reynolds number. This suggests a possible mechanism for the selection of the pattern amplitude. The most unstable wavelength scales as observed in nature on the thickness of the viscous sublayer, with a multiplicative factor that depends on the dimensionless parameters of the problem.
\end{abstract}

\begin{keywords}
Morphological instability, pattern formation, laminar-turbulent transition.
\end{keywords}

\section{Introduction}
\begin{figure}
\centerline{\includegraphics{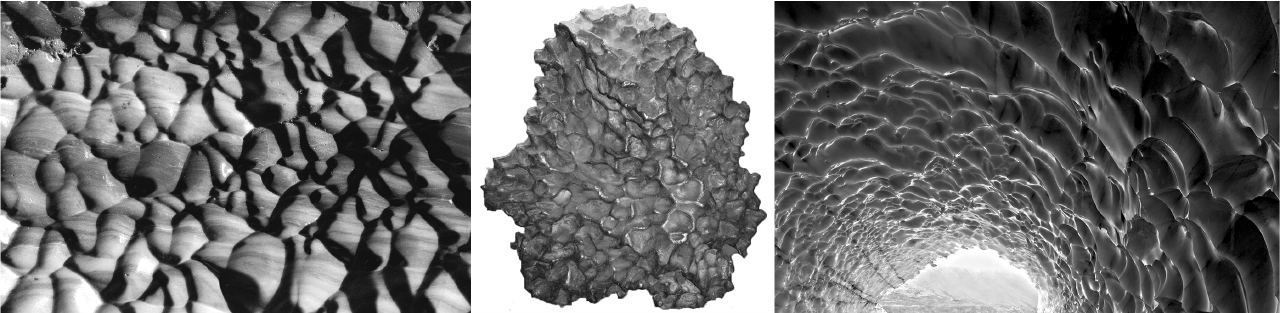}}
\caption{Pictures showing scallops on marble (left: Korallgrottan cave; photo width $\simeq 0.5$~m), at the surface of a meteorite (middle: Cabin Creek meteorite; width $\simeq 0.4$~m) and on ice (right: tunnel in Tarfala Valley; largest scallops $\simeq 1$~m).}
\vspace{-3 mm}
\label{Fig1}
\end{figure}

Pattern formation often occurs in nature when a fluid, flowing above a solid bed, can erode or transport some material of the bed. This mass transfer can be, for example, associated with sediment transport as for sand ripples and dunes (Charru et al., 2013). It can also be of thermodynamical origin with melting, sublimation or dissolution of the bed (Meakin \& Jamtveit, 2010), which this paper is devoted to. This occurs, for instance, in cave conduits, where the limestone dissolves in the water flow, forming scallops (Curl 1974; Blumberg \& Curl, 1974; Thomas, 1979). Similar scallop patterns form on meteorites with regmaglypts (Lin \& Qun, 1986; Claudin \& Ernstson, 2004), on glaciers and snow fields with suncups and ablation hollows (Rhodes et al., 1987; Herzfeld et al., 2003), at the surface of firn walls and caves and icebergs (Sharp, 1947; Leighly, 1948; Richardson \& Harper, 1957; Kiver \& Mumma, 1971; Anderson et al. 1998; Cohen et al., 2016), or in steel pipes under the action of corrosion (Lister et al., 1998, Villien et al., 2001). Water flows on ice can also produces ripples (Carey, 1966; Ashton \& Kennedy, 1972; Gilpin et al., 1980; Gilpin, 1981; Ogawa \& Furulawa, 2002), in particular on icicles (Ueno et al., 2010; Chen \& Morris, 2013). Other related studies concern the shape evolution of dissolving objects (Nakouzi et al., 2014; Huang et al., 2015), or the emergence of precipitation patterns in geothermal hot springs (Goldenfeld et al., 2006). A few examples are shown in Fig.~\ref{Fig1}.

In all of these situations, the flow is influenced by the bed elevation or profile, and in turn, erosion or transport induced by the flow makes the solid surface evolve. This feedback loop can lead to an instability, where bed perturbations are amplified. Several linear stability analyses have been performed for these dissolution/melting problems, in order to compute the growth rate of a perturbation of given wavenumber $k$ and determine the selected most unstable mode (Hanratty, 1981; Ogawa \& Furulawa, 2002; Ueno \& Farzaneh, 2011; Camporeale \& Ridolfi, 2012). In this paper, building on the pioneering work of Hanratty (1981), we incorporate the effect of the bed roughness in the hydrodynamical description, which is absent of previous analyses. This roughness turns out to be of key importance as the dissolution instability is found to disappear when the bed becomes rough; that is, above a threshold corresponding to a value of the roughness-based Reynolds number. We hypothesise that this restabilisation may explain the amplitude selection of the pattern.

The model proposed in this paper is composed of several parts. The first one, independent of the dissolution problem, deals with the description of a turbulent flow over a solid surface, using a turbulent closure relating the stresses to the velocity gradient (section~\ref{Hydrodynamics}). We use here a standard mixing length approach, with two specificities: we incorporate the role of the surface roughness and that of the pressure gradient on the turbulent mixing. Dissolution or melting is described by means of the advection-diffusion of a passive scalar (section~\ref{Scalar}) that can represent the temperature or the concentration of a dissolved species. It is coupled to the hydrodynamical part mainly because the coefficient of diffusion is related to the turbulent mixing. To address the development of an instability leading to the emergence of dissolution bedforms, we investigate this model in the case of a sinusoidally perturbed surface. The equations are linearised in the limit of a small perturbation and coupled to an erosion law for the bed evolution, to derive the dispersion relation of the problem (section~\ref{LSA}). It provides the growth rate and propagation velocity of the perturbation, and thus fully characterises the instability. Finally, the results are presented and discussed in section~\ref{Results}, and compared with experimental data.

\section{Reynolds averaged description and turbulent closure}
\label{Hydrodynamics}
We consider a fluid flow along the $x$ direction over a bed of elevation denoted by $Z$. Here, $z$ is the crosswise axis normal to the main bed, and $y$ is spanwise. Following the standard separation between average quantities and fluctuating ones (denoted by a prime), the equations governing the mean velocity field $u_i$ and the pressure $p$ can be written as
\begin{equation}
\partial_i u_i  = 0
\qquad {\rm and} \qquad
\rho \partial_t u_i+u_j \partial_j u_i  = \partial_j \tau_{ij}-\partial_i p,
\label{NS}
\end{equation}
where $\tau_{ij}$ contains the Reynolds stress tensor $-\rho \overline{u'_i u'_j}$. Here, $\rho$ is the density of the fluid, assumed to be constant. We use a first-order turbulence closure to relate the stress to the velocity gradient. It involves a turbulent viscosity resulting from the product of a mixing length and a mixing frequency, representing the typical eddy length and time scales (Pope, 2000). The mixing length $\ell$ depends explicitly on the distance from the bed. The mixing frequency is given by the strain rate modulus $|\dot \gamma| = \sqrt{\frac{1}{2} \dot \gamma_{ij} \dot \gamma_{ij}}$, where we have introduced the strain rate tensor $\dot \gamma_{ij}=\partial_i u_j+\partial_j u_i$. In a homogeneous situation along the $x$-axis, the strain rate reduces to $\partial_z u_x$. For a constant shear stress associated with a shear velocity $u_*$, we write
\begin{equation}
\tau_{xz} =\rho \ell^2 |\partial_z u_x| \partial_z u_x + \rho \nu \partial_z u_x=\rho |u_*|u_*,
\label{ConstantShearStress}
\end{equation}
where $\nu$ is the kinematic fluid viscosity. In order to account for both smooth and rough regimes, we adopt here a van Driest-like mixing length (Pope, 2000)
\begin{equation}
\ell=\kappa (z+rd-Z) \left[1-\exp\left(-\frac{(\tau_{xz}/\rho)^{1/2}(z+sd-Z)}{\nu \mathcal{R}_t}\right)\right].
\label{ellcombo}
\end{equation}
In this expression, $\kappa=0.4$ is the von K\'arm\'an constant, $d$ is the sand equivalent bed roughness size and $\mathcal{R}_t$ is the van Driest transitional Reynolds number, equal to $\mathcal{R}^0_t \simeq 25$ in the homogeneous case of a flat bed (Pope, 2000). The exponential term suppresses turbulent mixing within the viscous sub-layer, close enough to the bed $z=Z$. The dimensionless numbers $r=1/30$ and $s=1/3$ are calibrated with measurements of velocity profiles over varied rough walls (Schultz \& Flack, 2009; Flack \& Schultz, 2010). Here, $rd$ corresponds to the standard Prandtl hydrodynamical roughness extracted by extrapolating the logarithmic law of the wall at vanishing velocity. On the other hand, and this is more original, $sd$ controls the reduction of the viscous layer thickness upon increasing the bed roughness.

Following the work of Hanratty (1981), $\mathcal{R}_t$ cannot be taken as as constant but depends on a dimensionless number ${\mathcal H}$ that lags behind the pressure gradient following a relaxation equation:
\begin{equation}
a \frac{\nu}{u_*}\partial_x {\mathcal H}=\frac{\nu}{u_*^3} \partial_x (\tau_{xx}-p)-{\mathcal H},
\label{HanrattyRelaxation}
\end{equation}
where $a$ is the multiplicative factor in front of the space lag. We also introduce $b=\frac{1}{\mathcal{R}_t^0} \;\frac{d \mathcal{R}_t}{d {\mathcal H}}>0$ as the relative variation of $\mathcal{R}_t$ due to the pressure gradient. The values of these empirical parameters have been set to $a=2000$ and $b=35$, as calibrated on the behaviour of the basal shear stress over a modulated bed (Charru et al., 2013). Their values control the amplitude and location of the anomaly at the transition from a laminar to a turbulent response with respect to the bed elevation, as illustrated below.

We shall focus here on quasi two-dimensional situations, i.e. on geometries invariant along the $y$ direction. This simplification is a limitation for the study of non-transverse patterns, such as these scallops which have a kind of chevron shape, but is certainly enough to capture the physics of the instability. We can generally write the stress tensor as
\begin{equation}
\tau_{ij} = \rho \left ( \ell^2 |\dot \gamma| + \nu \right ) \dot \gamma_{ij} - \frac{1}{3} \rho \chi^2 \ell^2 |\dot \gamma|^2 \delta_{ij},
\label{relaxij}
\end{equation}
where $\delta_{ij}$ is the Kronecker symbol (see Fourri\`ere et al. (2010) and references therein). The typical value of the phenomenological constant $\chi$ is in the range $2$--$3$, but is of no importance here as we shall see that only the normal stress difference $\tau_{xx}-\tau_{zz}$ matters. This closed model allows us to compute the velocity and stress profiles for given boundary conditions, as illustrated in section~\ref{LSA}.

\begin{figure}
\centerline{\includegraphics{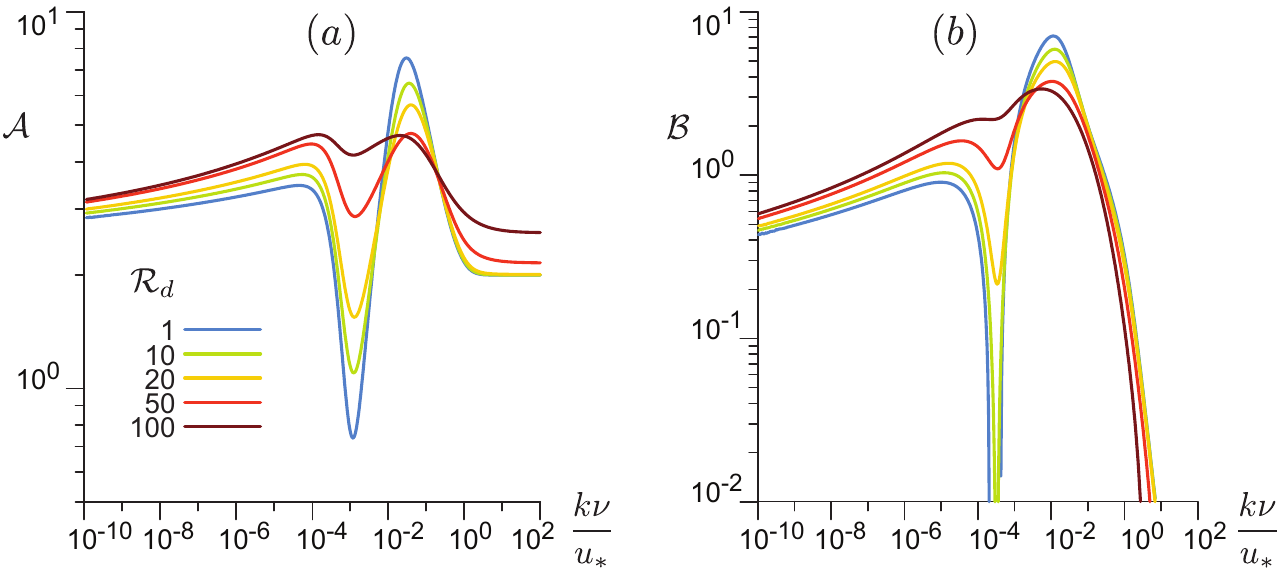}}
\caption{(a,b) Basal shear stress coefficients $\mathcal{A}$ and $\mathcal{B}$ computed by the model as functions of $k\nu/u_*$. The different colours code for the value of the parameter $\mathcal{R}_d$ (see legend). The laminar-turbulent transition (around $k\nu/u_* \simeq 10^{-3}$) gradually disappears when $\mathcal{R}_d$ increases.}
\label{Fig2}
\end{figure}

Before addressing the linear stability analysis, it is useful to discuss and show the effect of the bed roughness on the hydrodynamical quantities, here taking the shear stress as a typical example. Anticipating section~\ref{LSA} where a sinusoidally modulated bed $Z(x)=\zeta e^{ikx}$ of wavenumber $k$ is considered, these hydrodynamical equations can be linearised with respect to the small parameter $k\zeta$. In this case the shear stress is also modulated and correspondingly takes the generic form $\tau_{xz} = \rho u_*^2 (1+k\zeta e^{ikx} S_t) + {\rm c.c.}$, where $S_t(\eta)$ is a dimensionless function of the rescaled vertical coordinate $\eta=kz$. We define the two coefficients $\mathcal{A}$ and $\mathcal{B}$ by $S_t(0) = \mathcal{A} + i\mathcal{B}$: these are computed as outputs of the hydrodynamic model. The basal shear stress is then a sinusoidal function of $x$ and these coefficients encode the in-phase and in-quadrature components in response to the bed modulation. The coefficients $\mathcal{A}$ and $\mathcal{B}$ are functions of $k$, as displayed in Fig.~\ref{Fig2}. The main point, which is the hydrodynamical novelty of this study, is that they are also found to strongly depend on the bed roughness-based Reynolds number $\mathcal{R}_d = d u_*/\nu$. As described and discussed by Charru et al. (2013), these coefficients present, in the smooth limit ($\mathcal{R}_d <1$), a marked transition between the turbulent regime associated with small wavenumbers $k\nu/u_* \lesssim 10^{-4}$ and the laminar regime, typically for $k\nu/u_* \gtrsim 10^{-2}$. This transition, evidenced by the experimental data of Hanratty and his co-workers (Zilker et al., 1977; Frederick \& Hanratty, 1988), resembles a `crisis' with sharp variations of $\mathcal{A}$ and $\mathcal{B}$, allowing, in particular, for negative values of $\mathcal{B}$. The coefficients $a$ ad $b$ introduced above have been adjusted to fit these data (Charru et al., 2013). Crucially, this transition progressively disappears upon increasing the roughness Reynolds number, recovering the reference rough behaviour above $\mathcal{R}_d \gtrsim 100$ (Fourri\`ere et al., 2010). Beyond the phenomenology, a detailed physical understanding of this laminar-turbulent transition remains a pending hydrodynamical open problem.

\section{Scalar transport}
\label{Scalar}
We wish now to describe a passive scalar $\phi$, e.g. the concentration of a chemical species or the temperature, which is transported by the flow. We model its dynamics by a simple advection-diffusion equation,
\begin{equation}
\partial_t \phi+\dx q_x + \dz q_z = 0,
\label{AdvectionDiffusionEq}
\end{equation}
where the flux $\vec q$ is the sum of a convective and a diffusive term: $\vec q = \phi \vec u - D \vec \nabla \phi$. Here we take a diffusion coefficient proportional to the turbulent viscosity and write
\begin{equation}
D=\frac{\ell^2 |\dot{\gamma}|}{\beta_t}+\frac{\nu}{\beta_\nu},
\label{CoeffDiff}
\end{equation}
where $\beta_t$ and $\beta_\nu$ are the turbulent and viscous Schmidt numbers (or Prandtl numbers for temperature), here taken as constants. A typical value for liquids as well as gasses is $\beta_t=0.7$ (Gualtieri et al., 2017). For the molecular diffusivity, $\beta_\nu$ can be estimated from the Stokes-Einstein relation: $\frac{\nu}{\beta_\nu}=\frac{k_BT}{6\pi \rho \nu r_m}$, where $k_B$ is the Boltzmann constant, $T$ is the temperature, and $r_m$ is the molecular effective radius. The order of magnitude of molecular diffusion for ions or for dissolved CO$_2$ in water is $\beta_\nu=10^{3}$; for the diffusion of particles in an ideal gas, a typical value is $\beta_\nu=1$.

In the base state for which the bed is homogeneous in $x$, we can assume that the solid bed dissolves or melts at a small constant velocity and then perform the computation in the moving frame of reference. Eq.~\ref{AdvectionDiffusionEq} reduces to $\partial_z q_z = 0$, so that the flux is constant:
\begin{equation}
q_z=-D \dz \phi=q_0.
\label{qzbasestate}
\end{equation}
With the above expression (\ref{CoeffDiff}) for $D$, we obtain the equation for the vertical profile of the scalar field $\phi$,
\begin{equation}
\left(\frac{\ell^2 |\dot{\gamma}|}{\beta_t}+\frac{\nu}{\beta_\nu} \right) \partial_z \phi = -q_0.
\label{Profilephibasestate}
\end{equation}
To solve this equation, a condition on the bed must be specified. It depends on the nature of the scalar field, but can be formally written in a unique general way. For thermal problems, we impose the bed temperature: $\phi_0=T_0$. For dissolution or sublimation problems, we write a Hertz-Knudsen type of law (Eames et al., 1997), with a flux of matter that depends on the concentration at the surface,
\begin{equation}
q_0 = \alpha \left( \phi_{\rm sat} - \phi_0 \right).
\label{HertzKnudsen}
\end{equation}
Because this condition applies on the bed, where the velocity vanishes, no convective contribution to the flux is involved. In the above expression, $\phi_{\rm sat}$ is the concentration at saturation, and $\alpha$ is a reaction kinetic constant, here expressed as a velocity scale. For simplicity, we take it constant. In the case of sublimation or evaporation, $\alpha$ is given by the thermal velocity times a desorption probability (Eames et al., 1997), leading to $\alpha$ in the range $1$--$10$~m/s. In dissolution problems however, we expect $\alpha$ to take much smaller values, of the order of $10^{-5}$~m/s (Falter et al., 2005). Eq.~\ref{HertzKnudsen} can be re-expressed as $\phi_0 = \phi_{\rm sat} - q_0/\alpha$, so that in both cases $\phi_0$ is given. Finally, as buoyancy effects are usually negligible in such problems, we consider that the momentum equation is not coupled to the scalar field.

\section{Linearised equations}
\label{LSA}
We can now proceed to the linear stability analysis, to compute the dispersion relation.

\subsection{Definitions and base state}
For small enough amplitudes, we can consider a bottom profile of the form
\begin{equation}
Z(x)=\zeta e^{ikx}
\label{bottomprofile}
\end{equation}
without loss of generality (real parts of expressions are understood). Here, $\lambda=2\pi/k$ is the wavelength of the bottom and $\zeta$ the amplitude of the corrugation. The case of an arbitrary relief can be deduced by a simple superposition of Fourier modes. We introduce the dimensionless variable $\eta=k z$, $\eta_0=kd$ and the wavenumber-based Reynolds number ${\mathcal R}=\frac{ u_*}{k \nu}$. Primed quantities in this section denote derivatives with respect to $\eta$. With these notations and following (\ref{ellcombo}), the mixing length in the base state can be expressed in a dimensionless form as
\begin{equation}
k\ell \equiv \Upsilon(\eta)= \kappa (\eta+r \eta_0)\,(1-\exp(-{\mathcal R}(\eta+s\eta_0)/\mathcal{R}_t^0)).
\label{defUpsilon}
\end{equation}
We define the function $\mathcal{U}(\eta)$ giving the wind profile in the base state as $u_x \equiv u_* \mathcal{U}$. From (\ref{ConstantShearStress}) we see that it must verify the following equation:
\begin{equation}
\Upsilon^2 |\mathcal{U}'|\mathcal{U}'+ {\mathcal R}^{-1} \mathcal{U}'=1,
\qquad {\rm or\ equivalently} \qquad
\mathcal{U}'=\frac{-1 +\sqrt{1+4\Upsilon^2 {\mathcal R}^2}}{2\Upsilon^2 {\mathcal R}},
\label{EqDiffmathcalU}
\end{equation}
which must be solved with the boundary condition $\mathcal{U}(0)=0$ corresponding to the no-slip condition of the wind at the solid interface. Similarly, we define the function $\mathcal{P}(\eta)$ for the passive scalar in the base state as $\phi - \phi_0 \equiv q_0 \mathcal{P}/u_*$. From (\ref{Profilephibasestate}), it obeys
\begin{equation}
\left(\frac{\Upsilon^2 |\mathcal{U}'|}{\beta_t}+ \frac{{\mathcal R}^{-1} }{\beta_\nu}\, \right)\mathcal{P}'+1=0.
\label{EqDiffmathcalP}
\end{equation}
The boundary condition is $\mathcal{P}(0)=0$, corresponding to the condition $\phi = \phi_0$ on the bed.

\subsection{Linear expansion}
The aim of this subsection is to derive a set of closed equations (\ref{equaU'}-\ref{equaF'}), which, once integrated with given boundary conditions, provide as outputs the basal stresses, scalar concentration and flux, in order to compute the dispersion relation of the problem. Although technical, the principle of the calculation is simple and relies on the linearisation of the governing equations with respect to the small parameter $k\zeta$, which is proportional to the aspect ratio of the bed modulation. As in Fourri\`ere et al. (2010), all quantities are generically written as the sum of their homogenous term (along $x$) plus a linear correction generically written as $c k\zeta e^{ikx} C(\eta)$. The scaling factor $c$ encodes the physical dimension, whereas $C$ is a non-dimensional mode function which describes the vertical profile of the correction.
More explicitly, we write:
\begin{eqnarray}
u_x & = & u_* \left[\mathcal{U}+k\zeta e^{ikx} U \right],
\label{defU}\\
u_z & = & u_* k\zeta e^{ikx} W,
\label{defW}\\
\tau_{xz} & = & \tau_{zx}= \rho u_*^2 \left[1+k\zeta e^{ikx} S_t\right],
\label{defSt}\\
\tau_{zz}-p & = & -p_0 + \rho u_*^2 \left[ -\frac{1}{3}\chi^2 - k\zeta e^{ikx} S_n\right],
\label{defSn}\\
\tau_{zz} & = & \rho u_*^2  \left[ -\frac{1}{3}\chi^2 + k\zeta e^{ikx} S_{zz}\right],
\label{defSzz}\\
\tau_{xx} & = & \rho u_*^2  \left[- \frac{1}{3}\chi^2 + k\zeta e^{ikx} S_{xx}\right].
\label{defSxx}\\
k\ell & = &\Upsilon +  k\zeta e^{ikx} L,
\label{defL}\\
\phi & = &\phi_0 + \frac{q_0}{u_*}[\mathcal{P}+ k\zeta e^{ikx} \Phi],
\label{defmathcalP}\\
q_z & = & q_0 \left [1 - k\zeta e^{ikx} F \right],
\label{defF}
\end{eqnarray}
With these notations, we can express the shear rate as well as combination of these fields, e.g. for the computation of the coefficient of diffusion:
\begin{eqnarray}
 |\dot{\gamma}| & = & u_* k \left[\mathcal{U}'+(U'+iW)k\zeta e^{ikx} \right],
\label{dotgamma}\\
k \ell^2 |\dot{\gamma}|  & = & u_* \left[\mathcal{U}' \Upsilon^2+\left[(U'+iW)\Upsilon^2 + 2\Upsilon L \mathcal{U}' \right]k\zeta e^{ikx} \right],
\label{Reydcoe}\\
D & = & \frac{u_*}{k} \left[\left(\frac{\mathcal{U}' \Upsilon^2}{\beta_t} + \frac{\mathcal R^{-1}}{\beta_{\nu}}\right)+\frac{1}{\beta_t}\left[(U'+iW)\Upsilon^2 + 2\Upsilon L \mathcal{U}' \right]k\zeta e^{ikx} \right].
\label{coeffD}
\end{eqnarray}
For later use, we call $\Delta = \frac{1}{\beta_t}\left[(U'+iW)\Upsilon^2 + 2\Upsilon L \mathcal{U}' \right]$ the mode function of the coefficient of diffusion.

\begin{figure}
\centerline{\includegraphics{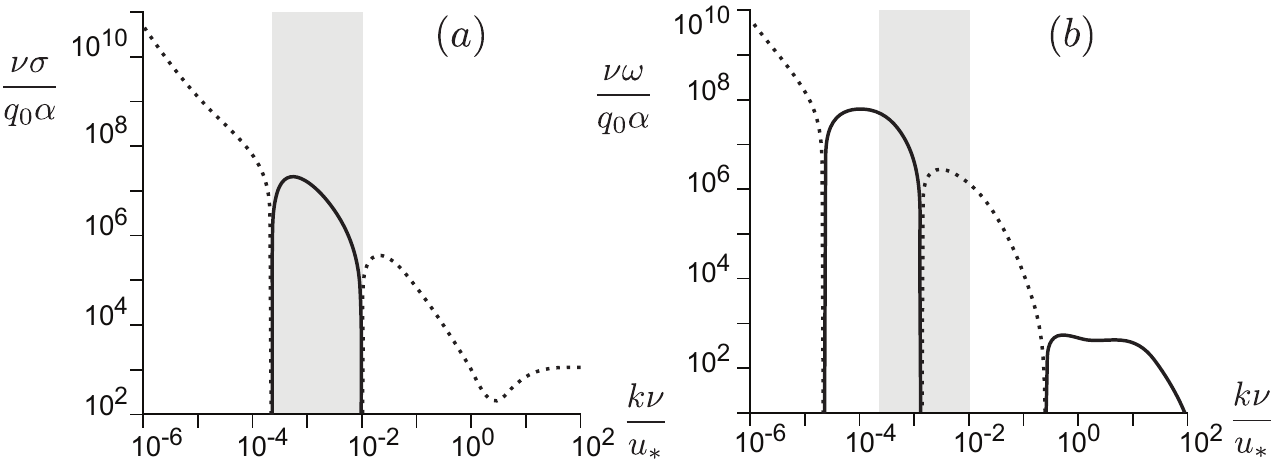}}
\caption{Dispersion relation for the dissolution instability in the limit of small $\alpha/u_*$, for $\beta_\nu=10^3$ and $\mathcal{R}_d = 10$. (a) Growth rate $\sigma$. (b) Angular frequency $\omega$. Solid lines represent positive values. For negative values, $-\sigma$ or $-\omega$ are plotted (dotted lines). The gray region corresponds to the unstable ($\sigma >0$) range of wavenumbers.}
\label{Fig3}
\end{figure}

Following (\ref{relaxij}), we express the stress functions as follows:
\begin{eqnarray}
S_{t} &=& \left( {\mathcal R}^{-1}+2 \Upsilon^2 \mathcal{U}' \right)(U'+iW)+2 \Upsilon \mathcal{U}'^2 L,
\label{Stexpress}\\
S_{xx}-S_{zz}  &=& 4\left( {\mathcal R}^{-1}+ \Upsilon^2 \mathcal{U}' \right) iU=\frac{4iU}{\mathcal{U}'},
\label{SssmoinsSzzexpress}
\end{eqnarray}
where we have used  $\left( {\mathcal R}^{-1}+ \Upsilon^2 \mathcal{U}' \right)=1/ \mathcal{U}'$ at the zeroth order (Eq.~\ref{EqDiffmathcalU}). From (\ref{ellcombo}), we can also express the disturbance to the mixing length:
\begin{equation}
L=\kappa \left \{-1+e^{- \frac{\mathcal R}{\mathcal{R}_t^0} (\eta+s\eta_0)} \left[1-\frac{\mathcal R}{\mathcal{R}_t^0} (\eta+r \eta_0)+\frac{\mathcal R}{\mathcal{R}_t^0} (\eta + s \eta_0)(\eta+r \eta_0) \left(\frac{1}{2}S_t - b {\mathcal H}\right)\right]\right\}.
\label{Lexpress}
\end{equation}
The linear expansion of the Navier-Stokes equations gives
\begin{eqnarray}
W' &=& -iU,
\label{NSmasslinear}\\
S_{t}' &=& i\mathcal{U} U +\mathcal{U}'W+iS_{n}-i S_{xx}+i S_{zz},
\label{NSxlinear}\\
S_{n}' &=& -i\mathcal{U} W + iS_{t}.
\label{NSzlinear}
\end{eqnarray}
Linearising the Hanratty equation (\ref{HanrattyRelaxation}), one obtains
\begin{equation}
\left({\mathcal R}+i a\right) {\mathcal H}=i(S_{xx}-S_{zz}-S_n)=-\frac{4U}{\mathcal{U}'}-iS_n.
\label{HanrattyLinearised}
\end{equation}
The linear expansion of the scalar equation gives
\begin{eqnarray}
F' &=& i{\mathcal P} U+\left(i\mathcal{U} +\frac{\Upsilon^2 |\mathcal{U}'| }{\beta_t}+ \frac{{\mathcal R}^{-1}}{\beta_\nu}\, \right)\Phi,
\label{Fprimeexpress}\\
{\rm with} \quad
F &=& \left(\frac{\Upsilon^2 |\mathcal{U}'| }{\beta_t}+ \frac{{\mathcal R}^{-1}}{\beta_\nu}\, \right)\Phi' + \frac{1}{\beta_t} {\mathcal P}'  \left[ \Upsilon^2 (U'+iW) +2 \Upsilon L\mathcal{U}'\right] -W{\mathcal P}.
\label{Fexpress}
\end{eqnarray}
Combining these equations, we finally obtain six closed equations
\begin{eqnarray}
U' &=& - i W +\frac{S_{t}-2  \Upsilon \mathcal{U}'^2 L}{{\mathcal R}^{-1}+2  \Upsilon^2 \mathcal{U}' },
\label{equaU'}\\
W' &=& -i U,
\label{equaW'}\\
S_t' &=& \left(i\mathcal{U} +\frac{4}{\mathcal{U}'}\right) U + \mathcal{U}' W + i S_n,
\label{equaSt'}\\
S_n' &=&-  i \mathcal{U}  W +i S_t,
\label{equaSn'}\\
\Phi' &=& \left[F+W{\mathcal P}-\frac{{\mathcal P}'   \left(\Upsilon^2 S_t+ 2 \Upsilon L\mathcal{U}' \left({\mathcal R}^{-1} +  \Upsilon^2 \mathcal{U}' \right)\right)}{\beta_t\left({\mathcal R}^{-1}+2  \Upsilon^2 \mathcal{U}' \right)}\right]/\left[\frac{\Upsilon^2 |\mathcal{U}'|}{\beta_t}+ \frac{{\mathcal R}^{-1}}{\beta_\nu}\right],
\label{equaPhi'}\\
F' &=& i{\mathcal P} U+\left(i\mathcal{U} +\frac{\Upsilon^2 |\mathcal{U}'| }{\beta_t}+ \frac{{\mathcal R}^{-1}}{\beta_\nu} \right)\Phi,\label{equaF'}
\end{eqnarray}
where the disturbance to the rescaled mixing length reads
\begin{eqnarray}
L=\kappa \left \{-1+e^{- \frac{\mathcal R}{\mathcal{R}_t^0} (\eta+s\eta_0)} \bigg[1 \right. &-& \frac{\mathcal R}{\mathcal{R}_t^0} (\eta+r \eta_0)
\label{Lexpress2}\\
&+& \left. \frac{\mathcal R}{\mathcal{R}_t^0} (\eta + s \eta_0)(\eta+r \eta_0) \left(\frac{1}{2}S_t + \frac{b}{{\mathcal R}+ia} \left( \frac{4U}{\mathcal{U}'}+iS_n \right)\right)\bigg]\right\}.
\nonumber
\end{eqnarray}
%

\begin{figure}
\centerline{\includegraphics{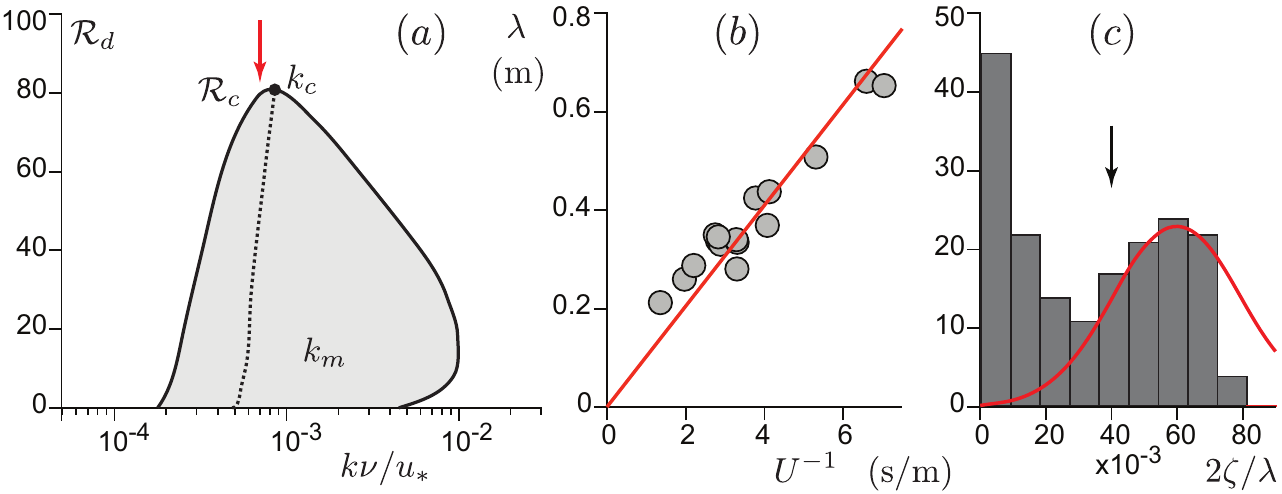}}
\caption{(a) Stability diagram in the plane $(\mathcal{R}_d, k\nu/u_*)$, computed in the limit of small $\alpha/u_*$ and for $\beta_\nu=10^3$. Solid line: marginal stability curve ($\sigma=0$). Grey: unstable zone ($\sigma>0$). Dotted line: location of the most unstable modes ($\sigma_m$). (b) Experimental data of Ashton \& Kennedy (1972) on ice ripples: Mean wavelength \emph{vs} inverse of mean flow velocity. The linear fit $\lambda \simeq 0.1/U$ (solid line) corresponds to the red arrow in panel a. (c) Histogram of the ripple aspect ratio in these experiments. The black arrow corresponds to the aspect ratio giving ${\mathcal R}_d={\mathcal R}_c$}
\label{Fig4}
\end{figure}

\subsection{Boundary conditions}
Six boundary conditions must be specified to solve the system (\ref{equaU'}-\ref{equaF'}), labelled (i-vi) below. The upper boundary corresponds to the limit $\eta \to \infty$, in which the vertical fluxes of mass and momentum vanish asymptotically. This means that the first-order corrections to the shear stress and to the vertical velocity must tend to zero: (i) $W(\infty)=0$ and (ii) $S_t(\infty)=0$. In practice, we introduce a finite height $H$ (or $\eta_H \equiv kH$), at which we impose a null vertical velocity and a constant tangential stress $-\rho u_*^2$, so that $W(\eta_H) = 0$ and $S_t(\eta_H) = 0$. Then, we consider the limit $H \to +\infty$, i.e. when the results become independent of $H$. On the bed $z = Z$, both velocity components must vanish, which gives: (iii) $U(0) = - \mathcal{U}'(0)$ and (iv) $W(0) = 0$. Using (\ref{EqDiffmathcalU}) and (\ref{defUpsilon}), we can express
\begin{equation}
U(0) =  \frac{1 - \sqrt{1+4\Upsilon(0)^2 {\mathcal R}^2}}{2\Upsilon(0)^2 {\mathcal R}}
= \frac{1 - \sqrt{1+4 [\kappa r \eta_0 \left( 1 - \exp(-s\eta_0 {\mathcal R}/\mathcal{R}_t^0) \right)]^2 {\mathcal R}^2}}{2[\kappa r \eta_0 \left( 1 - \exp(-s\eta_0 {\mathcal R}/\mathcal{R}_t^0) \right)]^2 {\mathcal R}} \, .
\label{Uof0bis}
\end{equation}
Regarding the passive scalar, we hypothesise that its flux through the upper boundary remains constant, which gives: (v) $F(\eta_H)=0$. Finally, on the bed, the dissolution-like condition $q_z = \alpha (\phi_{\rm sat} - \phi)$, associated with its zeroth order (\ref{HertzKnudsen}), leads to: (vi) $F(0) = \frac{\alpha}{u_*} \left[ \mathcal{P}'(0) + \Phi(0) \right]$, where $\mathcal{P}'(0)$ is known from (\ref{EqDiffmathcalP}). Other quantities like $S_t(0)$ and $S_n(0)$ come as results of the calculation.

\begin{figure}
\centerline{\includegraphics{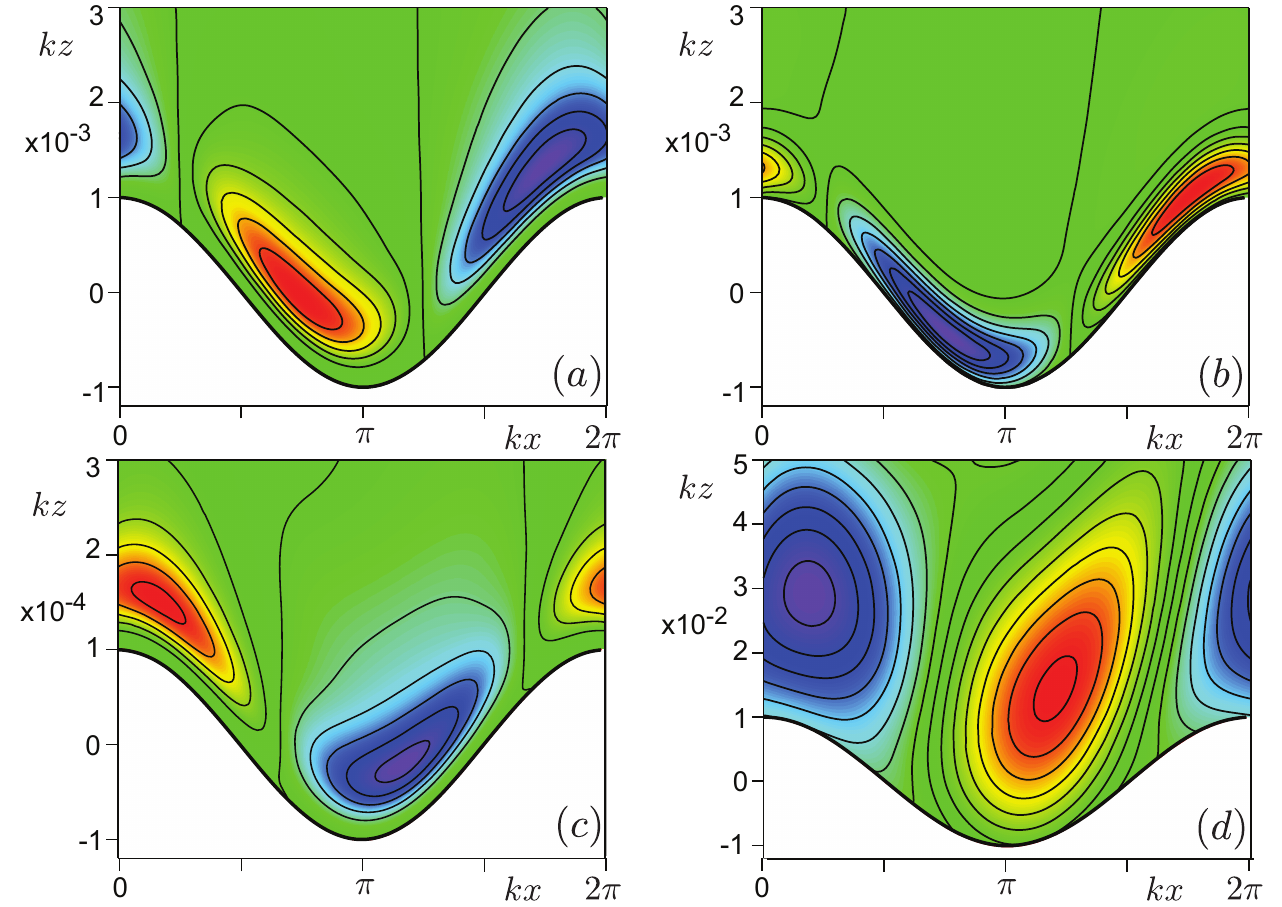}}
\caption{(a) Isocontours of the diffusion coefficient above a modulated bed (in white), with a wavenumber close to the most unstable mode: $k\nu/u_*=5 \cdot 10^{-4}$. The other parameter values are $\mathcal{R}_d = 0.1$, $\alpha/u_*=0.01$, $\eta_H=10$ and $\beta_\nu=10^3$. Red (blue) regions correspond to a strong (weak) mixing. (b) Corresponding isocontours of the concentration $\phi$. Strong mixing leads to a large vertical flux which reduces $\phi$. (c) Same as (a), but for a smaller wavenumber $k\nu/u_*=5 \cdot 10^{-5}$. (d) Isocontours of the concentration $\phi$ at a larger wavenumber: $k\nu/u_*=5 \cdot 10^{-2}$. In all panels, the flow is from left to right.}
\label{Fig5}
\end{figure}

As both $F(0)$ and $\phi(0)$ must remain bounded to be compatible with the bulk equations, we expect two simple asymptotic regimes: in the limit of small $\alpha/u_*$, the scalar modulation $\Phi(0)$ results from the mixing of the base profile and is independent of $\alpha$; the flux modulation $F(0)$ follows and must vanish linearly in $\alpha/u_*$. Conversely, at large $\alpha/u_*$ the situation is opposite: the substrate is so erodible that any disturbance in concentration at the surface would lead to a diverging flux resorbing it. As $\Phi(0)$ vanishes as $u_*/\alpha$, the flux $F(0)$ results from the mixing of the base state and is therefore constant.

\subsection{Interface growth rate}
To compute the temporal evolution of the bed elevation, we proceed with $Z(x,t) = \zeta e^{\sigma t +i\omega t+ ikx}$, where $\sigma(k)$ is the growth rate and $\omega(k)$ is the angular frequency of the bed pattern along $x$. The phase propagation speed is therefore $-\omega/k$ with these notations. As the equations are linear in $\phi$, one can always define the relevant scalar with the appropriate factor so that the evolution equation for the bottom reads $\partial_t Z = q_0 - q_z(Z)$. At the linear order, this gives the dispersion relation,
\begin{equation}
\sigma + i\omega = q_0 k F(0).
\label{DispersionRelation}
\end{equation}
%

\section{Results and discussion}
\label{Results}
The dispersion relation (\ref{DispersionRelation}) is displayed in Fig.~\ref{Fig3}, in the limit of small $\alpha/u_*$ and at vanishing $\mathcal{R}_d$ (smooth case). Following the asymptotic behaviour of $F(0)$ in this limit, the relevant rescaling factor for $\sigma$ is $q_0 \alpha/\nu$. We see in panel (a) a range of unstable wavenumbers with a positive growth rate, in which $\sigma$ reaches a maximum value $\sigma_m$ at $k_m$. In this range, the propagation velocity changes sign (panel b), showing that the instability is absolute and not convective. As shown in Fig.~\ref{Fig4}a, the key result is that the unstable band disappears above a critical value $\mathcal{R}_c$ of the roughness Reynolds number.

\begin{figure}
\centerline{\includegraphics{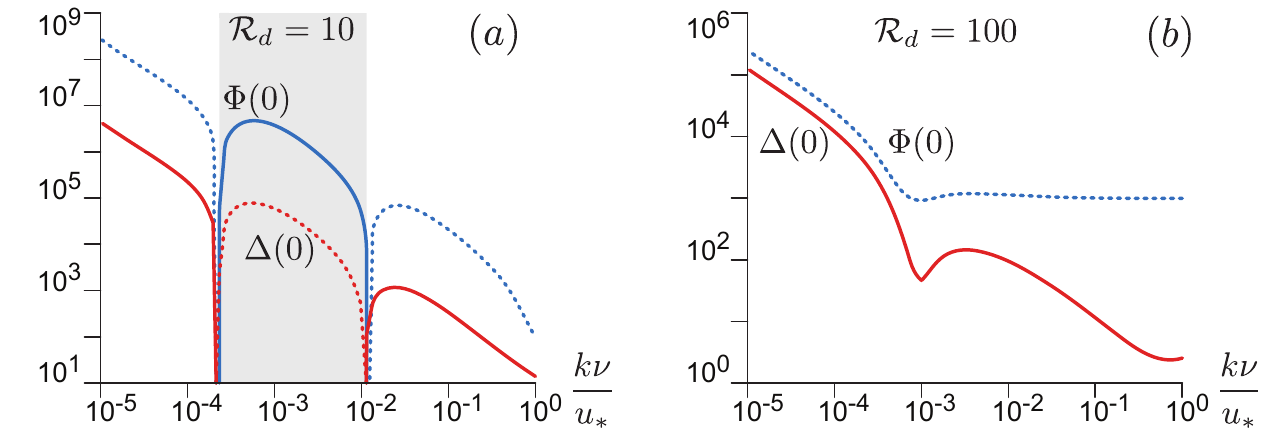}}
\caption{In-phase components of the modulation of the diffusion coefficient $\Delta(0)$ in red and of the basal concentration $\Phi(0)$ in blue, for $\mathcal{R}_d=10$ (a) and $\mathcal{R}_d=100$ (b), as functions of $k\nu/u_*$. Dashed lines represent negative values. The gray region is the unstable range of wavenumbers.}
\label{Fig6}
\end{figure}

As illustrated in Fig.~\ref{Fig5}, one can understand the instability mechanism as follows. The erosion of the bed is driven by the mass flux $q_z$, itself controlled by the concentration gradient and the coefficient of diffusion (Eq.~\ref{qzbasestate}). The concentration profile, enforced by the base state, is non-homogeneous, decreasing away from the surface. The crests of a modulated bed profile come closer to regions of lower concentration, enhancing the gradient with respect to the surface where $\phi$ is imposed. For a constant $D$, this peak effect increases the flux and thus the erosion at the crests, and this stabilising situation is what happens at large $k\nu/u_*$, when the wavelength is much smaller than the viscous-sublayer. When turbulence is dominant, $D$ is not constant any more, but is controlled by turbulent mixing. At small $k\nu/u_*$, turbulence is enhanced slightly up-stream of the crests, and hence there is stabilising erosion again. For wavenumbers in the intermediate range corresponding to the laminar-turbulent transition, however, turbulence is shifted downstream by means of the adverse pressure gradient (Eq.~\ref{HanrattyRelaxation}), enhancing mixing and thus erosion in the troughs, which a is destabilising (amplifying) situation. The opposite behaviour of the in-phase modulations of $D$ and $\phi$ is displayed in Fig.~\ref{Fig6} for the whole range of wavenumbers, showing a change of sign corresponding to enhanced mixing in troughs in the presence of the laminar-turbulent transition only.

\begin{figure}
\centerline{\includegraphics{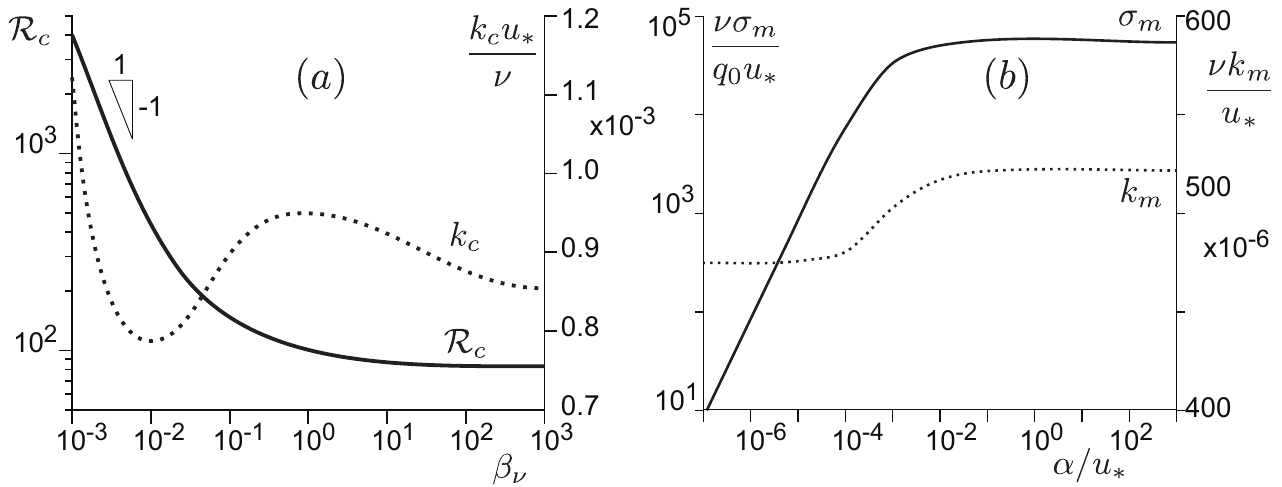}}
\caption{(a) Critical value $\mathcal{R}_c$, at which the instability disappears, as a function of $\beta_\nu$ (solid line, left axis), in the limit of small $\alpha/u_*$. Right axis, dotted line: corresponding critical wavenumber $k_c$. (b) Maximum growth rate (solid line, left axis), and corresponding wavenumber (dotted line, right axis) as functions of $\alpha/u_*$, computed for $\beta_\nu=10^3$ and $\mathcal{R}_d\to 0$.}
\label{Fig7}
\end{figure}

Changing the molecular diffusivity, this phenomenology with a range of unstable modes (Fig.~\ref{Fig4}a) remains, but with consequently varying values of $\mathcal{R}_c$ and $k_c$ (Fig~\ref{Fig7}a). For $\mathcal{R}_c$ we clearly identify a low-$\beta_\nu$ regime where $\mathcal{R}_c \propto 1/\beta_\nu$ and a plateau at large $\beta_\nu$. This means that diffusive processes are controlled by whatever is dominant between the diffusion of momentum ($\nu$) and that of dissolved species ($\nu/\beta_\nu$). Moreover, $k_c$ is found to be pretty constant, independent of $\beta_\nu$, with a typical value of around $10^{-3} u_*/\nu$. This relative invariance is what one can expect for a bandpass instability. Exploring a broad range of the ratio $\alpha/u_*$ in Fig.~\ref{Fig7}b, but now rescaling the growth rate by $q_0 u_*/\nu$, we see that $\sigma_m$ becomes independent of it, with a crossover at around $\alpha/u_* \simeq 10^{-3}$. These regimes correspond to those discussed for $F(0)$, in relation to (\ref{DispersionRelation}). Meanwhile, the most unstable wavenumber $k_m$ switches from a plateau value to another value, with a small relative change, emphasising again that relevant wavenumbers are fairly insensitive to all parameters.

The development of the instability actually increases the bed roughness, suggesting that the pattern eventually selects nonlinearly the wavenumber $k_c$. As a matter of fact, converting the amplitude of the bed perturbation to a sand equivalent roughness size $d$ (Flack \& Schultz, 2010), the value $\mathcal{R}_d \simeq 80$ is reached with a pattern aspect ratio $2\zeta/\lambda$ of the order of $5$\%, for these values of $k$, i.e. typically $k\zeta \simeq 0.16$. This value is clearly the upper bound for the linear expansion to make sense, but still reasonably small.

Thomas (1979) has gathered measurements from various experiments that provide evidence of the global scaling of the selected scallop wavenumber with the viscous length. The fit of these data gives $k \simeq 6 \cdot 10^{-3} u_*/\nu$. This is in fair agreement with our results, but to be more quantitative, we must emphasise two difficulties when looking in more detail at specific experiments. A first ambiguity lies in the definition of the wavelength for three-dimensional objects like scallops. For example, in the study of a dissolution pattern on plaster by Blumberg \& Curl (1974), $\lambda$ is identified as ratio $\left< \Lambda^3 \right>/\left< \Lambda^2 \right>$, where the angle brackets denote average over measurements of longitudinal scallop sizes $\Lambda$. With this definition, a selected wavenumber $k \simeq 3 \cdot 10^{-3} u_*/\nu$ is reported. Another issue in computing the value of $k\nu/u_*$ from experimental data is the difficulty of relying on an accurate estimate of the shear velocity, as also discussed by Blumberg \& Curl (1974). Usually, the mean flow velocity is actually measured, and $u_*$ is computed assuming a velocity profile, typically the logarithmic law of the wall, which leads to $u_* \simeq \kappa U/\ln \frac{H}{z_0}$, where $H$ is the flow depth and $z_0$ is the hydrodynamical roughness, itself related to the bedform amplitude as $z_0=rd$.

Closer to the two-dimensional situation that we consider here, we have more quantitatively investigated the data of Ashton \& Kennedy (1972) for ice ripples. These authors report a clear linear law $\lambda \sim 1/U$ (Fig.~\ref{Fig4}b). They have also measured the ripple aspect ratio. This quantity is widely distributed (Fig.~\ref{Fig4}c), showing a population of emerging bedforms with small aspect ratios, and another one of mature ripples, with an aspect ratio centred around $6$\% (Fig.~\ref{Fig4}c). Computing $u_*$ from $U$ as discussed above, we obtain for these data $k\nu/u_* \simeq 8 \cdot 10^{-4}$, in quantitative agreement with the value of $k_c$ (Fig.~\ref{Fig4}a). Further experimental studies are needed to investigate the emergence and development of this instability in more detail, and in particular to follow the evolution of the bed roughness over time (Villien et al., 2005). Another direction of research is the investigation of fast flows in order to connect scallops generated by water flows with those on meteorites, for which supersonic effects are expected.

\acknowledgments
We thank F. Charru and G. Vignoles for discussions and L. Tuckerman for a critical reading of the manuscript.


\end{document}